\documentclass[fleqn,12pt,twoside]{article}
\usepackage{espcrc1}
\usepackage{fleqn}

\usepackage{graphicx}
\usepackage{subfigure}
\usepackage[figuresright]{rotating}
\usepackage[latin1]{inputenc}      

\newcommand{\AmS}{{\protect\the\textfont2
  A\kern-.1667em\lower.5ex\hbox{M}\kern-.125emS}}

\newcommand{\AM}{$A\;$MeV}

\title{Correlations between signals of the liquid-gas phase transition in
nuclei}

\author{M. F. Rivet\address[IPNO]{Institut de Physique Nucléaire, F-91406
Orsay cedex, France},
        N. Le Neindre\addressmark,
        J. P. Wieleczko\address[GANIL]{GANIL, F-14076 Caen, France},
	B. Borderie\addressmark[IPNO],
	R. Bougault\address[LPC]{LPC, ENSICAEN et Université, F-14050 Caen, France},
	A. Chbihi\addressmark[GANIL],
	J. D. Frankland\addressmark[GANIL],
        M. Pârlog\address[IFIN]{NIPNE, RO-76900 Bucharest-M\u{a}gurele, 
	Romania}%
	\thanks{Present address: GANIL, F-14076 Caen, France},
        M. Pichon\addressmark[LPC],
	B. Tamain\addressmark[LPC]
        and INDRA and ALADIN Collaborations
}       
\begin{document}

\maketitle

\begin{abstract}
Finite systems such as atomic nuclei present at phase transition specific 
features different from those observed at the thermodynamic limit.
Several characteristic signals were found in samples of events resulting 
from heavy ion collisions at and above the Fermi
energy. The concomitant observation of different signatures of a liquid-gas 
phase transition  in nuclei on a given sample strongly supports the 
occurrence of this transition.
\end{abstract}

\section{INTRODUCTION}
A liquid-gas type phase transition is theoretically expected for nuclei
owing to the analogy between the nucleon-nucleon interaction and the van 
der Waals force. Experimentally, indications of phase transition have been 
reported many years ago by groups studying multifragmentation. In the 
last few years progresses developed in the two domains: statistical physics 
built the concepts associated to the definition and signatures of phase 
transition in finite systems, among which atomic nuclei; on the experimental 
side new and performant 4$\pi$ multidetectors allowed to study carefully 
selected samples of events instead of working on singles data. 
Among those the INDRA array was used to detect charged products formed in a
wide variety of nuclear collisions, which bring nuclei 
in a state where phase transition will occur. The quality of the exclusive
measurements permitted to search for different signatures of the transition, 
as it seems obvious that the concomitant observation of several of them 
will strongly reinforce the hypothesis that a phase transition has occurred.  

\section{EXPERIMENTAL EVENT SAMPLES}

The results presented in this contribution were obtained in experiments
performed with the 4$\pi$ array INDRA operating at GANIL or at GSI and mostly 
concern systems, or sources, with about 200-230 nucleons: ``fused
sources'' from Xe+Sn, Ni+Au reactions, Au or Ta quasi-projectiles. In order
to analyse data in the framework of statistical physics, one must select
well defined sources. In central collisions this was done by two techniques:
either a selection of compact sources by taking events whose main direction
of emission deviates by  more than 60$^o$ from the beam direction 
(Xe+Sn system)~\cite{I28-Fra01}; or for the asymmetrical Ni+Au system 
a discriminant analysis method~\cite{I37-Bel02}. 
In the case of ``fused sources'' the experimental distributions
of excitation energy are narrow.

The selection is more delicate when one wants to isolate quasi-projectiles,
because of the presence of direct and neck emissions. For symmetric systems,
a first order separation is to identify as quasi-projectile (quasi-target) 
the ensemble of products with velocities parallel to the beam direction
larger (smaller) than the centre of mass velocity. 
This method was used for the Au+Au data discussed later. The advantage of
quasi-projectiles is that the whole possible range of excitation energy
(0 - available energy) can be explored.

In all cases a tolerance of about 10\% on the source charge is imposed.

Any of these data samples constitutes a statistical ensemble, in the sense
of a collection of events with similar properties. To apply statistical 
physics analyses one must be careful about the correct framework
(microcanonical, canonical ensembles, with or without dynamical constraints
{\ldots} ).

\section{SIGNALS OF PHASE TRANSITION : THEORETICAL BASES}\label{theo}
While at the thermodynamic limit first order phase transitions are
characterized by
discontinuities or divergences of observables, smoother situations are
expected in finite systems, which exhibit anomalous curvatures in
thermodynamical potentials. Saying that the entropy, S(E), presents a convex
curvature at first order phase transition is fully equivalent to stating
that the energy distribution is bimodal, meaning that at a given temperature
correspond two different values of the energy.
Another consequence is that, in a
microcanonical framework, the caloric curve, T(E), should present a
backbending for the energy domain corresponding to the phase transition,
and therefore the heat capacity, $c$=dE/dT, will have a negative
branch. However caloric curves may differ depending on the thermodynamical
path followed by the system in the temperature-pressure-energy space,
and may not present any backbending, even at phase transition.
A more robust signal was found in
the fluctuations of the kinetic part of the energy, directly connected
to the heat capacity: very large fluctuations sign a negative value of
$c$. See~\cite{Cho02,Gro02} for reviews.

In this picture, one then expects to observe at the same time a 
bimodality of the energy and a negative heat capacity as signals of a first
order phase transition in nuclei. Moreover as the system is then located 
in the mechanically unstable spinodal region of the 
phase diagram, additional information can be found 
in searching for characterization of the dynamics of the phase transition, 
which in nuclei may proceed through nucleation or spinodal decomposition. 
Many theoretical predictions being in favour of the latter~\cite{Cho04}, 
a third signal 
of phase transition might be the evidence of spinodal decomposition.

On another hand the first indications of phase transition in the nineties
were related to critical behaviors, such as the observation of a power law
mass distribution for light products of multifragmentation~\cite{MorIWM01}. 
At the thermodynamic limit, scalings are general properties of matter near
the critical point and thus typical of second order phase transitions. 
Recent studies have however shown that scaling
properties were observed on a much larger range when the considered system
is finite: it is sufficient that the correlation 
length be close to the system size to observe scalings. 
In a lattice gas framework, for instance, it was demonstrated that
Fisher scaling occurs when the system lies on a ``critical line'' which
extends well inside the coexistence region~\cite{Gul99}. 
\emph{For nuclei scalings can
therefore also be considered as indications of the occurrence of a first 
order phase transition.}
 
\section{SIGNALS OF PHASE TRANSITION : EXPERIMENTAL EVIDENCES}\label{exp}

Most of the data mentioned in this section are published, we will not
illustrate them here but rather send the reader to the 
quoted references.

\subsection{Dynamics of the phase transition: spinodal
decomposition}\label{SD}

Theoretical studies as well as semi-classical simulations of collisions 
showed that, in the course of the reaction, the system is driven to the
spinodal region of the phase diagram~\cite{Jac96,Gua97}. If low enough 
densities are reached, the system will never recover to normal conditions, 
but rather break-up into pieces. The rupture is initiated by local density 
fluctuations which are amplified in an homogeneous way by the mean field, 
creating regular high density regions which will finally turn into equal 
size fragments, surrounded by nucleons and very small fragments (the gas). 
The signature of spinodal decomposition in infinite nuclear matter is thus 
the observation in the exit channel of fragments of a given size, linked to 
the wave length of the most unstable mode. Here again the picture is different 
for nuclei, because of their finiteness. Firstly in this case several 
modes have equal characteristic times, leading to fluctuations in the size 
and multiplicity of the fragments from event to event~\cite{Jac96}. 
Beating of modes and coalescence during fragment separation may 
break the symmetry of each partition.  And last, as fragments are  born 
hot, the 
de-excitation stage also alters the partitions. As a result the favoured
partitions resulting from spinodal decomposition of nuclei might only survive
with a very small probability, making them difficult to reveal.

The method proposed to evidence enhanced equal-size fragment partitions
uses higher order charge correlations~\cite{Mor96}, each event being 
characterized by the average and the standard deviation of its fragment 
(Z$\geq$5) charge distribution. 
The delicate point in any correlation function is 
the construction of uncorrelated events. The initial method did not account
for charge conservation; the INDRA collaboration proposed two methods 
to remedy this drawback, which lead to different results. 
 Applied to semi-classical simulations of collisions between Xe and Sn at 
32\AM, where it is known that spinodal decomposition does occur, 
the modified event mixing method did not show any print of favoured 
partitions~\cite{I41-Cha03}, while the intrinsic probability method (IPM) 
enlightens an excess of about 1\% of events where fragments have equal 
sizes. On experimental data the results of the two methods also differ, 
the event mixing method always leading to less positive conclusions. 
With the IPM, evidence of spinodal decomposition was
found, at the level of 0.3-1\% of the events, in central collisions
between Xe and Sn at 32, 39 and 45\AM, the signal vanishing at
50\AM~\cite{I40-Tab03}. 
It appeared at 52\AM{} for central Ni+Au collisions, but not at the
lower energy of 32\AM~\cite{T34Gui02}. The thermal energy put in the systems
showing prints of spinodal decomposition was evaluated (with the help 
of a statistical model) to 5-7\AM~\cite{RivIWM01}.

\subsection{Bimodality}\label{Bimod}
 By definition, at first order phase transition, the distribution of an
order parameter, or of any related variable, should be bimodal in a
canonical framework. This was looked for in Au quasi-projectiles from
Au+Au collisions between 60 and 100\AM{}.
To mimic a canonical, or a least a gaussian
ensemble~\cite{Cha88}, the sorting for the quasi-projectile was performed 
as a function of the transverse energy of the light charged particles
(lcp, Z$\leq 2$) emitted backward of the centre of mass, $E_{t12}^{QT}$.
This is justified by the high efficiency of INDRA for lcp in the entire
phase space.
At first order $E_{t12}^{QT}$  quantifies the dissipated energy.
For a specific bin of $E_{t12}^{QT}$ a bimodal distribution is found 
for the charge asymmetry of the two largest fragments of each 
event, $(Z_{max} - Z_{max-1}) / (Z_{max} + Z_{max-1}) $, when plotted versus
$Z_{max}$, see examples in fig.~\ref{QPAu+a} and~\ref{QPAu+b}: two groups
of events clearly stand out, large asymmetries correspond to a big fragment
associated with one or several very small ones (residue, or liquid type), 
while asymmetries close to 0 are associated with events with more and 
smaller fragments (gas type). 
Outside this specific  region, charge asymmetries are close to 1 for 
low $E_{t12}^{QT}$,
and turn to 0 for high $E_{t12}^{QT}$~\cite{TamEPS04}.

In the transition $E_{t12}^{QT}$ region, the properties of the two groups 
of events were investigated. The small asymmetry events have larger
multiplicities of both fragments and lcp, indicating
that the excitation energy of these events is larger than that of the other
group. This is confirmed by
calorimetry measurements: the energy distributions associated with the two
groups of events are distinct, although overlapping~\cite{TamEPS04}. 
If one admits
that a selection of $E_{t12}^{QT}$ fixes the temperature, then two values of
the energy are correlated to it, which is the signature of a bimodal behaviour.

\subsection{Negative heat capacities}\label{cneg}
The microcanonical heat capacity is a thermodynamic variable which has
the specific property to
present a negative branch in the transition zone for finite systems.
Experimental determinations of the heat capacity rely on the measurements of
kinetic energy fluctuations at freeze-out~\cite{Cho02}.
The heat capacity is given by the relation 
$c = c_{k}^{2}/(c_{k}-(<A_{0}> \times \sigma_{k}^{2})/T^{2})$,
where $c_{k} = \delta <E_{k}/A_{0}> / \delta T$ is the kinetic heat capacity,
$A_0$ and $T$ the mass and
temperature of the system and $\sigma_{k}$ the kinetic energy fluctuation.
The heat capacity is negative when the fluctuation is larger than the
reference kinetic heat capacity.

The method suffers however several drawbacks, 
some from lacks in the measured quantities (number and
energy of neutrons, fragment masses), and others intrinsic to the method
(reconstruction of the freeze-out configuration from the measured products).
The influence of the implied hypotheses was carefully tested on
event samples obtained in a statistical model, SMM~\cite{MDA02}. 
Interestingly it was demonstrated that the method tends to decrease the 
fluctuations, meaning that large fluctuations, and thus 
negative heat capacities, are not an artefact created by the imperfections
of the measurement.

First observed on Au quasi-projectiles from 35\AM{} Au+Au semi-peripheral
collisions~\cite{MDA99}, negative heat capacities were also seen in INDRA
data for the same reactions at 60-100A~MeV incident energies~\cite{T38Pic04}.
The results of both experiments are in good agreement.
In central collisions the end of the negative branch and the recovering to
positive values
were seen in Xe+Sn~\cite{T25NLN99} and Ni+Au~\cite{T34Gui02} collisions.
The ranges of excitation energies where negative values of $c$ were measured
are in good enough agreement in all these cases, if one reminds that in
central collisions some collective energy has to be added to the potential
energy part.

\subsection{Scalings}\label{sca}

Two types of scalings were studied in the INDRA data. Fisher
scaling was already found in many experiments, and critical exponents were
extracted. $\Delta$-scaling was proposed more recently for determining the
scenario of a second order phase transition. 

\subsubsection{Fisher scaling}\label{Fishc}
\begin{figure}[htb]
\subfigure[Excitation functions for different nuclei
produced in 39\AM{} Ta+Au semi-peripheral collisions. Symbols are
experimental data, lines show a fit with the Fisher formula.]{%
\label{Fisher+a}%
\begin{minipage}{0.45\textwidth}
\includegraphics[scale=0.78]{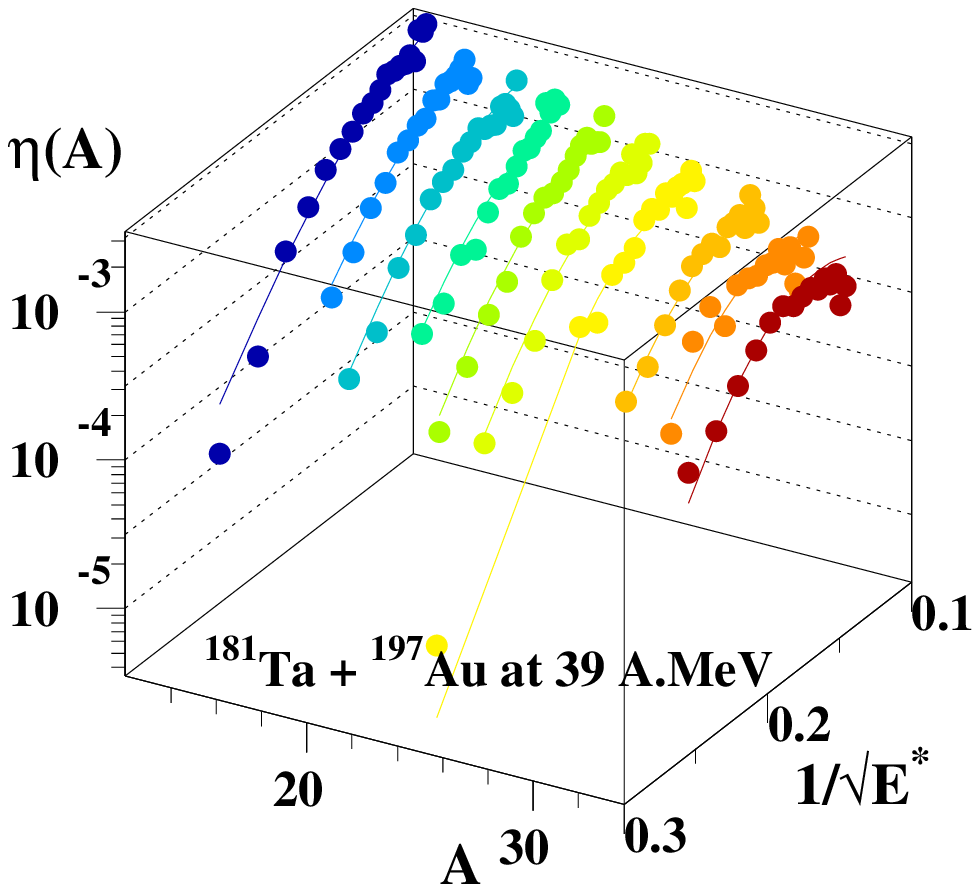}
\end{minipage}}
\subfigure[Scaled
cross sections for nuclei from boron to phosphorus produced in
semi-peripheral (Ta+X) or central (Xe+Sn, Ni+Au) collisions. Points for the
different system, which all lie on a single curve, are shifted for a
better view.]{%
\label{Fisher+b}%
\begin{minipage}{0.55\textwidth}
\centering \includegraphics[scale=0.78]{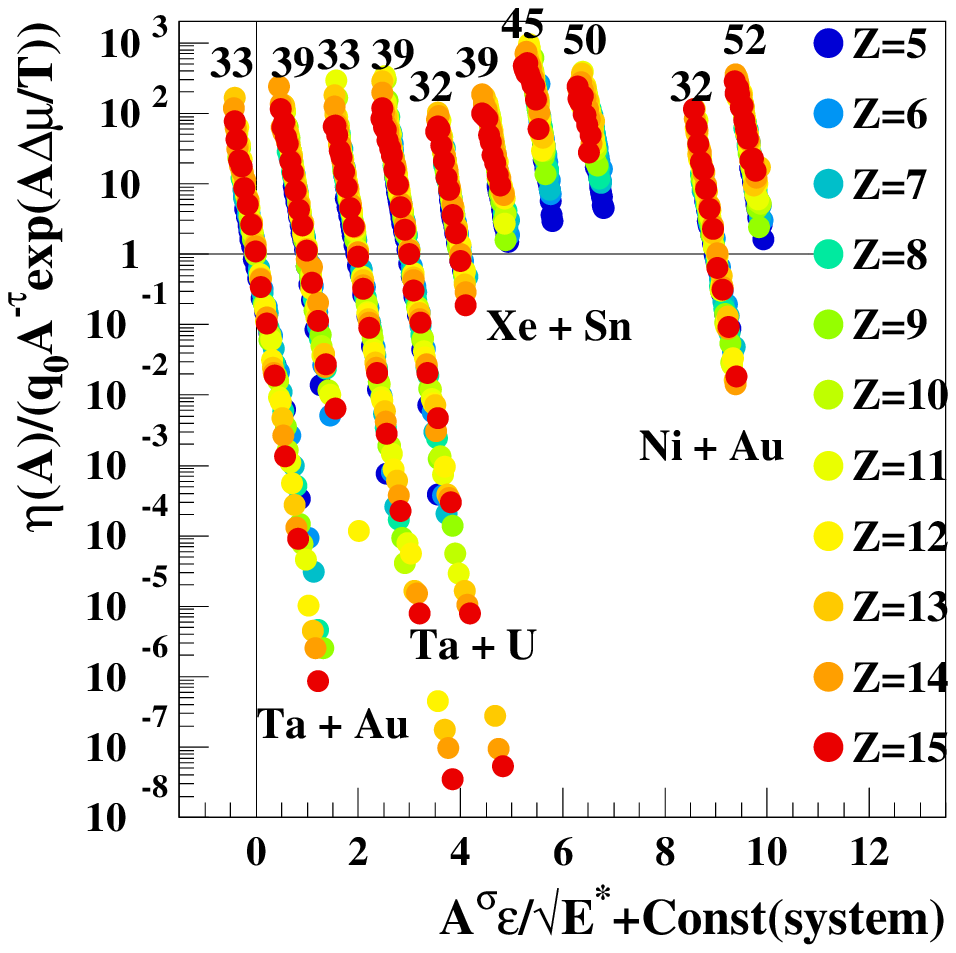}
\end{minipage}}
\vspace*{-1cm}
\caption{Fisher scaling } \label{Fisher}
\end{figure}
Fisher droplet model was originally proposed to study the morphology of
fluids in thermal equilibrium in terms of an ideal gas of clusters
coexisting with a liquid fraction. This
model was recently applied to multifragmentation data, by considering all
fragments but the largest as the gas phase, Z$_{max}$ being assimilated
to the liquid part. The yield of a fragment of mass A reads:
 $\mathrm{d} N / \mathrm{d} A = \eta(A) = 
 q_{0}A^{-\tau} \exp((A\Delta\mu-c_{0}\varepsilon A^{\sigma})/T)$.
In this expression, $\tau$ and $\sigma$ are critical exponents, $\Delta \mu$
is the difference between the chemical potentials of the two phases,
$c_0$ is the surface energy coefficient;
$\varepsilon = (T_{c}-T)/T_{c}$ describes the distance of the
actual to the critical temperature. This kind of scaling was found in many
multifragmentation data, in hadron-nucleus as well as in nucleus-nucleus
collisions; the agreement between data and theory often holds over orders of
magnitude, and the critical exponents which are deduced are in good
agreement with those expected for the liquid-gas universality class.
In INDRA data, Fisher scaling was applied to fragments obtained in single
sources as well as in quasi-projectiles~\cite{NLNBorm02}.
An example of the excitation functions for light fragments emitted 
by tantalum quasi-projectiles is given in fig.~\ref{Fisher+a};
the Fisher formula well fits the experimental points over large ranges of
masses and temperatures. Note that here the
temperature is replaced by the square-root of the excitation energy,
measured by calorimetry, according to the Fermi gas relation $E^*=aT^2$. 
In fig.~\ref{Fisher+b} are shown the scaling properties of the data obtained 
for different fused systems (Xe+Sn, Ni+Au) or Ta
quasi-projectiles from four different reactions; once scaled, all data
collapse onto a single curve -  they have been shifted along the abscissa
on the figure for better visualisation. 
For quasi-projectiles, and for fused sources formed at 32\AM, the scaling
extends below and above the critical temperature. 
This unexpected behaviour comes from
the finite size of the systems. As shown in~\cite{Gul99}, for finite systems
the ``Fisher critical temperature''  obtained  is not the true critical one 
if the system is in the coexistence region, and it 
varies with the density of the system. It happens for the data shown in
fig.~\ref{Fisher+b}
that the critical excitation energies found are in all cases 
close to 4.5\AM~\cite{NLNBorm02}. If one remembers that 
the systems depicted in the figure also have similar masses, and thus may be
represented by a single phase diagram, the unicity of the critical energy
might be an indication that the density at phase transition 
(when the system multifragments) is similar for all
systems, be they fused systems formed in violent collisions or 
quasi-projectiles from more peripheral collisions.

The values found for the topological exponent $\tau$ (2.1-2.4) 
and the one related to the surface-volume dimensionality ratio 
$\sigma$ (0.66) are in agreement
with previous data, close to the ones of the liquid-gas universality class.

\subsubsection{Universal scaling}\label{Delta}

 The theory of universal fluctuations of order parameters  provides
 information on the relationship between the phase transition of a system
 and the formation of clusters, without  requiring a precise knowledge of
 the thermodynamical state of this system~\cite{Bot00}. All information is
 supposed to be contained in the multiplicity and size of the clusters. For
 systems which exhibit second-order critical behaviours, the critical order
parameters can be identified through their $\Delta$-scaling behaviour.

In the INDRA multifragmentation data on central collisions between a large
variety of heavy ions,
it appeared that the fragment multiplicity did not present any scaling,
at variance with the size of the heaviest fragment, Z$_{max}$. The nature of
the order parameter identifies multifragmentation as an aggregation 
scenario~\cite{Ixx-Fra04}. Symmetric
systems with total masses between 73 and 394 all showed two scaling regimes,
from $\Delta$=1/2 at low incident energy to $\Delta$=1 at higher
energies, indicating the passage from an ordered phase to a disordered
phase; the transition energy decreases when the systems grow heavier.
(see~\cite{Ixx-Fra04,FraEPS04} for details). It is interesting to note that
the scaling behaviour is robust; for central collisions, it is rather 
independent of the precise event selection: the choice of compact 
sources~\cite{Bot01} leads to Z$_{max}$
distributions which are slightly narrower and shifted towards lower values
than those obtained with an impact parameter selector~\cite{Ixx-Fra04}, but
the transition energy is barely displaced.

$\Delta$-scaling was also evidenced in quasi-projectiles, provided that a
selection is made via the bimodality order parameter. Events with large
asymmetries scale with $\Delta$=1/2 whereas those with more equal fragments
scale with $\Delta$=1, see fig.~\ref{QPAu+d}. 

The $\Delta$-scaling signature of a phase transition is up to now the only
one which has been observed without ambiguity for systems of total mass
lighter than 150.

\section{SIMULTANEOUS OBSERVATION OF SEVERAL SIGNALS}

As said above, the hypothesis that atomic nuclei undergo a first order 
phase transition is strongly strengthened if several signatures of the
transition  are found on a given sample of events. Two examples will be 
presented in this section.

\subsection{Quasi-projectiles from Au+Au reactions}
\begin{figure}[tb]
\subfigure[Transverse energy distribution of lcp measured backward of 
the c.m. velocity, normalised to the incident energy per nucleon. 
The bars show the region where bimodality occurs.]{%
\label{QPAu+a} %
\begin{minipage}[t]{0.5\textwidth}
\includegraphics[width=7cm,height=6cm]{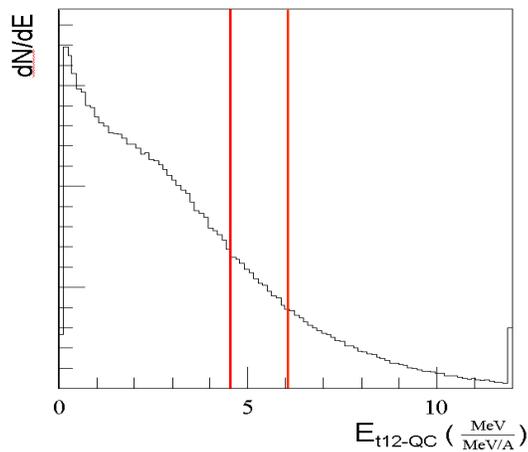}
\end{minipage}}
\subfigure[Charge of the largest fragment vs charge asymmetry of the two
largest fragments, Z$_{asym}$.]{%
\label{QPAu+b} %
\begin{minipage}[t]{0.5\textwidth}
\includegraphics[width=8cm,height=6cm]{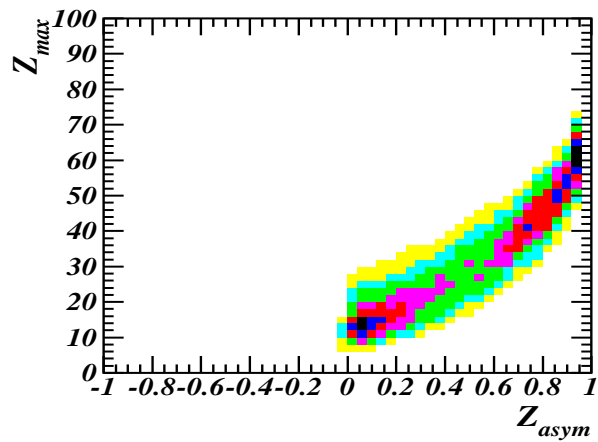}
\end{minipage}} 
\subfigure[Kinetic heat capacity (stars, dotted line) and kinetic energy 
fluctuations (other symbols, full line.) Lines are to guide the eye.]{%
\label{QPAu+c} %
\begin{minipage}[t]{0.5\textwidth}
\includegraphics*[width=8cm,height=6cm]{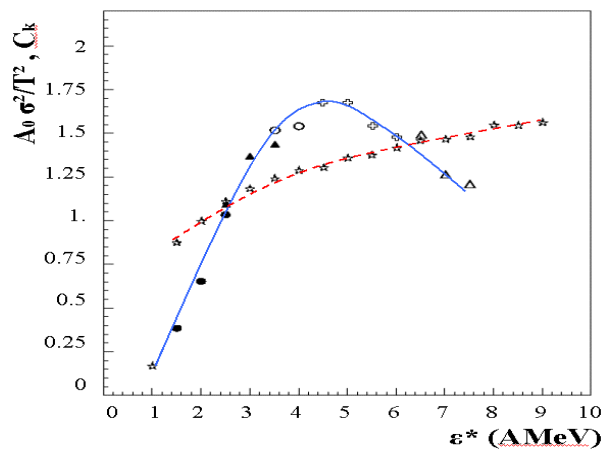}
\end{minipage}}
\subfigure[$\Delta$-scaling. Events with large charge asymmetry lie on the
right branch ($\Delta$=1/2), those with small asymmetry are on the left branch
($\Delta$=1).]{%
\label{QPAu+d} %
\begin{minipage}[t]{0.5\textwidth}
\includegraphics*[width=8cm,height=6.5cm]{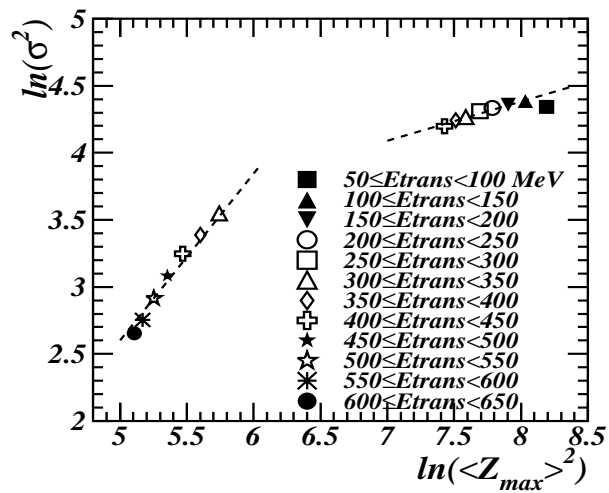}
\end{minipage}}
\vspace*{-1cm}
\caption{Quasi-projectiles from 80\AM{} Au on Au collisions. Part (b) is
obtained for events inside the bars in (a). 
See~\cite{TamEPS04} for details. (from~\cite{T38Pic04})
 } \label{QPAu} 
\end{figure}
The first case, shown in fig~\ref{QPAu},
is that of quasi-projectiles (QP) produced in Au+Au reactions 
at 80\AM{} measured with INDRA at GSI. As explained in~\ref{Bimod} the
properties of quasi-projectiles were followed after sorting on
$E_{t12}^{QT}$, the distribution of which is shown in fig.~\ref{QPAu+a}.
Bimodality neatly occurs in the delimited $E_{t12}^{QT}$ region, as shown in
fig.~\ref{QPAu+b}.

The $\Delta$-scaling was searched for in all $E_{t12}^{QT}$ bins, and
appears when separating the two groups of events.
Events associated with large asymmetries (or $Z_{max}$) belong to the
$\Delta$=1/2 family, while those with small asymmetries correspond
to $\Delta$=1: the transition is found in the
energy region where bimodality occurs (diamonds and crosses in
fig.~\ref{QPAu+d}).
Finally the negative heat capacity was calculated in a microcanonical
framework inside each $E_{t12}^{QT}$ bin.
A further selection on event compactness was added for this
study~\cite{T38Pic04}. The excitation energy of quasi-projectiles is
determined by calorimetry. In figure~\ref{QPAu+c} are shown the kinetic energy
fluctuations (solid line), 
versus the QP energy. Different symbols correspond to different bins of
$E_{t12}^{QT}$. Fluctuations  overcome the kinetic heat capacity (dotted line)
in a small excitation energy domain of the QP (3-6\AM), signing a negative
value of the heat capacity; the largest
fluctuations are obtained for events inside the $E_{t12}^{QT}$ bin where
bimodality occurs (crosses in fig.~\ref{QPAu+c}). Note that
fluctuations are maximum for an excitation energy of $\sim$4.5A MeV, which
is also the Fisher critical energy for similar (Ta) QP, as indicated in the
previous section.

Finally, for the selected sample of Au quasi-projectiles
excited at about 4-5\AM{}  two signals characteristic of a first order phase
transition are found at the same time. The observation of a scaling law
also supports the existence of such a transition, because the system studied
is finite.

\subsection{Single-source events from Xe+Sn reactions}
\begin{figure}[tbp]
\subfigure[Excitation function for the spinodal signal between 32 and 50\AM{}
(from~\cite{I40-Tab03}).]{%
\label{XeSn+a} %
\begin{minipage}[c]{0.37\textwidth}
\includegraphics*[scale=0.6]{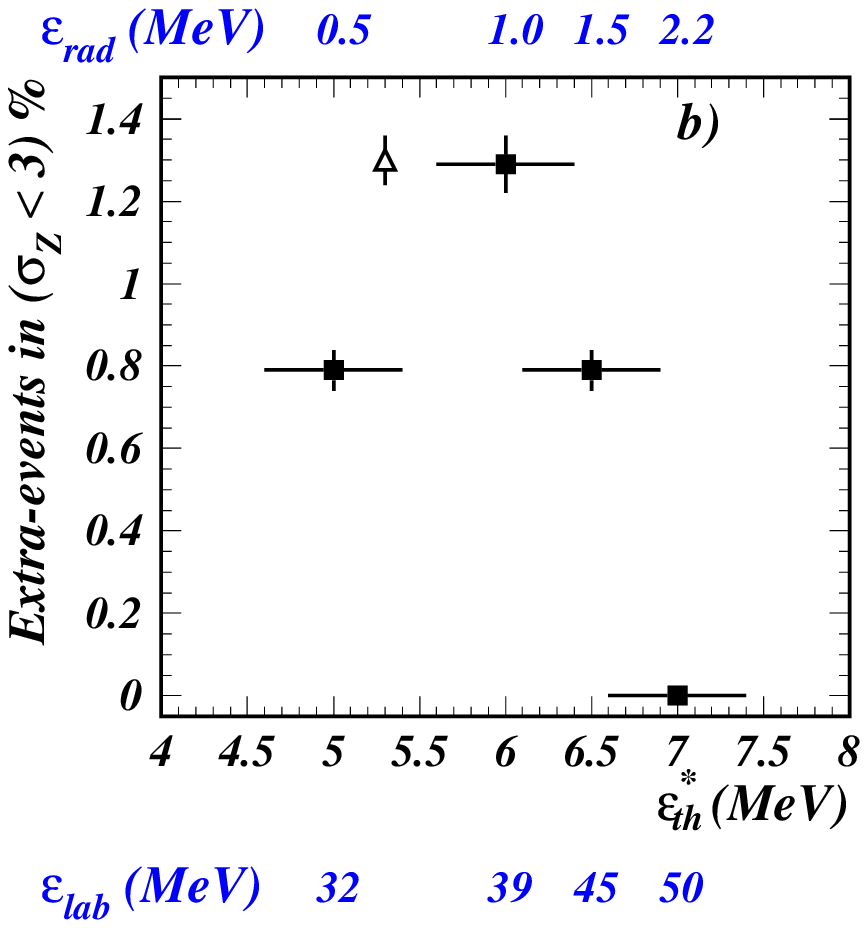}
\end{minipage}}
\subfigure[Energy distribution and heat capacity for collsions at 32 and
39\AM{}  (from~\cite{MDA02}).]{%
\label{XeSn+b} %
\begin{minipage}[c]{0.6\textwidth}
\includegraphics*{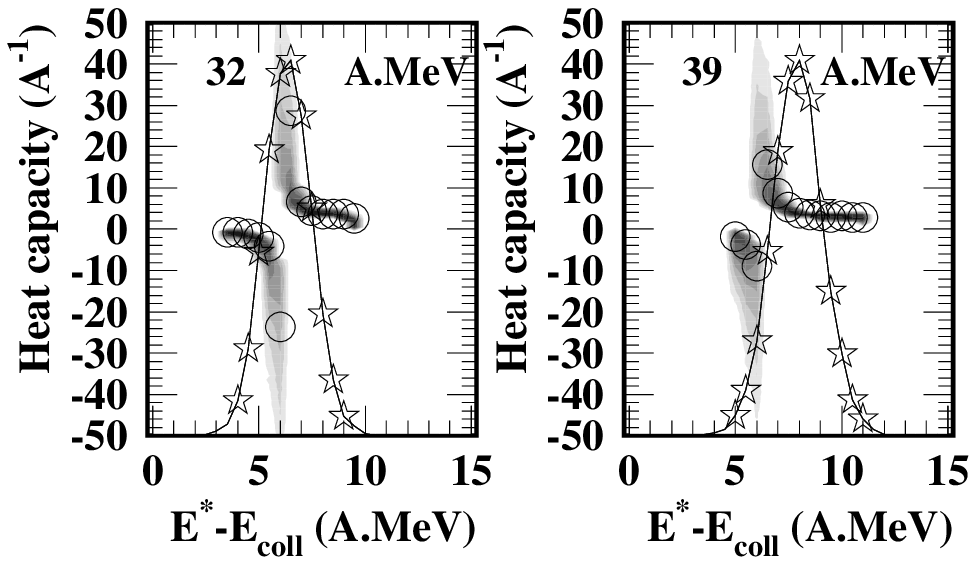}%
\end{minipage}}\\%
\subfigure[Fisher scaling at 32, 39, 45 and 50\AM (from~\cite{NLNBorm02}).]{%
\label{XeSn+c}%
\begin{minipage}[t]{0.4\textwidth}
 \includegraphics*[scale=0.9]{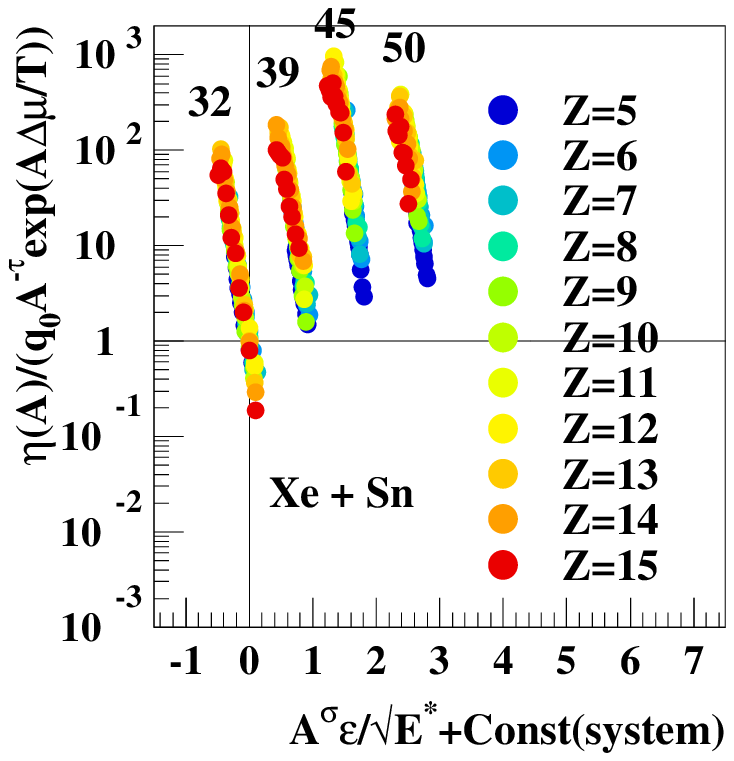}
\end{minipage}}
\subfigure[$\Delta$-scaling. Points from right to left were obtained for
collisions at 25, 32, 39, 45 and 50\AM. (from~\cite{Bot01}).]{%
\label{XeSn+d}%
\begin{minipage}[t]{0.6\textwidth}
\centering \includegraphics*[scale=1.1]{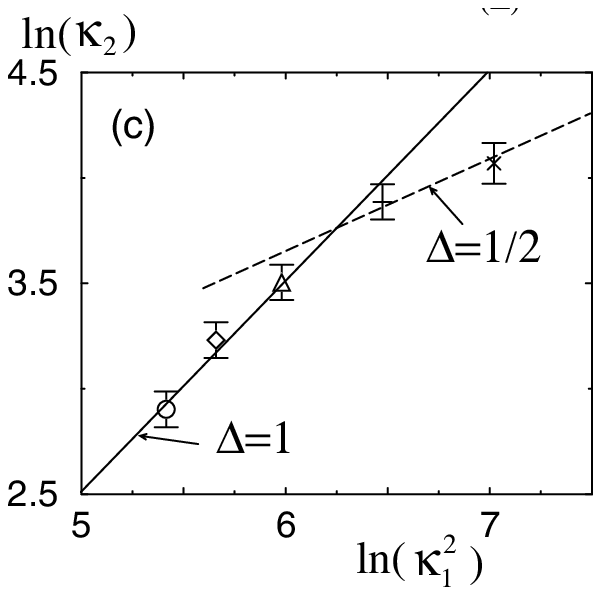}
\end{minipage}}
\vspace*{-1cm}
\caption{Central collisions between Xe and Sn from 25 to 50\AM}\label{XeSn}
\end{figure}
The second example, displayed in fig~\ref{XeSn},
concerns fused sources formed in central xenon on tin
collisions at 25, 32, 39, 45 and 50\AM. The event samples are in all cases
obtained with the same selection (compact event shapes). Fig~\ref{XeSn+a}
shows the excitation function of the excess of events with equal-sized 
fragments,
signature of a spinodal decomposition. The trace of the phenomenon is
maximum at 39\AM. The end of the negative branch and the restoring 
to a positive value of the heat capacity are
seen for the 32 and 39\AM{} samples (fig.~\ref{XeSn+b}). From the Fisher
scaling one finds a critical excitation energy of 4.5\AM, explored in the
reaction at 32\AM{} only; one expects a change in the multifragmentation
pattern around this energy, which matches the energy where the 
kinetic energy fluctuations are maximum ~\cite{NLNBorm02}. 
Finally the transition from
order to disorder marked by the change in $\Delta$  scaling also occurs 
between 32 and 39\AM (fig.~\ref{XeSn+d}). 

In summary, in events from central Xe on Sn collisions  we have evidenced four
signals indicating that the phase transition region is explored for incident
energies between 32 and 39\AM. The transition energies found in the different
analyses do not fully coincide; calorimetry gives about 6\AM, while a
determination through a comparison with the SMM model indicates 5\AM,
close to the Fisher critical energy of 4.5\AM. 
All values are however in agreement within the present precision
of about 1\AM{} that we have on calorimetry measurements.

\section{Summary and prospects}

The two above examples clearly indicate that one can characterize a
first order phase transition in nuclei with masses around 200, independently
of the violence of the collision in which they were formed, and of the
amount of radial expansion energy (larger in central than in peripheral
collisions). Work is in progress to look for signatures of
phase transition on lighter systems.
To be more quantitative and do metrology, refinements of the
methods are in progress, and experiments with larger statistical samples are
necessary. In the future more information will be obtained when new
accelerators will furnish very exotic beams, permitting to explore the
influence of the N/Z degree of freedom on phase transition. Calculations
predict for instance the shrinking of the spinodal region when the number
of neutron becomes very large~\cite{Col02}. Ultimately, reliable information should be
gained on the symmetry term of the equation of state.


\end{document}